\documentclass[aps,prl,twocolumn,showpacs,preprintnumbers]{revtex4}
\usepackage{graphicx}
\usepackage{amssymb}
\usepackage{amsmath}
\usepackage{times}
\usepackage{latexsym}
\def\beq{\begin{equation}}
\def\eeq{\end{equation}}
\def\bey{\begin{eqnarray}}
\def\eey{\end{eqnarray}}

\def\lsim{\mathrel{\raise.3ex\hbox{$<$\kern-.75em\lower1ex\hbox{$\sim$}}}}
\def\gsim{\mathrel{\raise.3ex\hbox{$>$\kern-.75em\lower1ex\hbox{$\sim$}}}}

\newcommand{\be}{\begin{equation}}
\newcommand{\ee}{\end{equation}}

\newcommand{\gev}{{\rm ~GeV }}

\begin{document}

\preprint{MCTP/11-03}

\title{A Theory of Maximal Flavor Violation}
\author{Jessie Shelton$^ a$ and Kathryn M. Zurek$^ b $}
\address{$^a$ Department of Physics, Yale University, New Haven, CT 06511\\
$^b$ Michigan Center for Theoretical Physics, University of Michigan, Ann Arbor, MI 48109  }

\date{\today}

\begin{abstract}

Theories of flavor violation beyond the Standard Model typically suppose that the new contributions must be small, for example suppressed by Yukawa couplings, as in Minimal Flavor Violation.  We show that this need not be true, presenting a case for flavor violation which maximally mixes the first and third generation flavors.  As an example, we realize this scenario via new right-handed gauge bosons, which couple predominantly to the combinations $(u,b)_R$ and $(t,d)_R$.  We show that this new flavor violation could be responsible for several anomalies, focusing in particular on the $B_d,~B_s$ systems, and the Tevatron top forward-backward asymmetry.


\end{abstract}
\maketitle


A battery of tests from LEP, the Tevatron and $B$ factories have proven the Standard Model (SM) to be a robust model for gauge interactions and flavor physics.  Because of the success of the SM, the conventional wisdom is that new physics Beyond the SM should approximately respect flavor, with new flavor violation suppressed by the Yukawa couplings, as in Minimal Flavor Violation.  Our understanding is based on measurements of the CKM unitarity triangle, which so far seem to hold up quite well. 

At the same time, there are a number of observations that suggest the possibility of new flavor violating physics.  The CDF experiment has observed a large forward-backward asymmetry in $t\bar{t}$ production: $A^{t\bar{t}} = 0.475 \pm 0.114$
for $M_{t\bar{t}} > 450 \mbox{ GeV}$, which is $3.4 \sigma$ discrepant from the leading order QCD prediction for the asymmetry, $A^{t\bar{t}} _{SM}= 0.088 \pm 0.013$ \cite{Aaltonen:2011kc}.  D0 also observes an anomalously large asymmetry, but with lower statistical significance \cite{Abazov:2007qb}.
A class of models which generates the asymmetry makes use of flavor violation in the $t$-channel, taking $u \rightarrow t$ or $d \rightarrow t$ through $t$-channel $Z'$ \cite{Jung}, $W'$ \cite{BargerTopAFB,WagnerTopAFB} or di-quarks \cite{TaitAFB}.

In addition, several $B$ physics measurements show discrepancies with the SM that hint at new flavor violation \cite{Lenz:2010gu,Lunghi:2010gv}.  The  D0 collaboration measured the like-sign dimuon charge asymmetry in semileptonic $b$ decays to be $a_{SL}^b = -(9.57 \pm 2.51 \pm 1.46) \times 10^{-3}$,  $3.2\sigma$ away from the SM prediction, $a_{SL}^b = -2.3^{+0.5}_{-0.6}\times 10^{-4}$ \cite{Abazov:2010hv}.  CDF's measurement of this quantity is consistent with the SM, but with much larger errors: $a_{sl}^b = 8.0 \pm 9.0\pm 6.8)\times 10^{-3}$ \cite{CDFdimuon}.  Combining the two results in quadrature gives $a_{sl}^b = -(8.5 \pm 2.8) \times 10^{-3}$, still a $3 \sigma$ deviation from the SM.  To further strengthen the case for new flavor physics, the measurements of $\Delta \Gamma_s$ and $S_{\psi \phi}$ made by both D0 \cite{Abazov:2008fj} and CDF \cite{CDFpsiphi} reported some deviation in $B_s-\bar{B}_s$ mixing from the SM; the magnitude and sign of the deviation are consistent with the deviation in $a_{SL}^b$ observed by D0 \cite{Bona:2008jn}.  Note that the Tevatron $a_{SL}^b$ measurement does not distinguish between $B_s$ and $B_d$ mesons, so that in principle,  both $B_s-\bar{B}_s$ and $B_d-\bar{B}_d$ mixing could contribute.  Other (less significant) indications for new CP violation in $B_d$ mixing and decays include the measurement of $\sin 2\beta$ in the tree-level decay $B_d \rightarrow \psi K_s$ \cite{Lunghi:2010gv,Ligeti} and in the penguin-dominated $b\rightarrow s \bar{q} q$ transitions of $B_d \rightarrow (\phi,\eta',\pi,\rho,\omega) K_s$ \cite{bargerFNUZp}.

The purpose of this paper is to write a theory with maximal flavor violation (MaxFV) which may be responsible for generating some or all of these anomalies.  The specific phenomenological model we have in mind is a new $SU(2)_R$ gauge boson that couples to flavor in the combinations $(u,b)_R$ and $(t,d)_R$.  As has already been demonstrated, the $(t,d)_R$ coupling alone can generate the Tevatron forward-backward asymmetry for an appropriate choice of the couplings and gauge boson masses, as can a flavor-violating $Z'$ coupling $u_R$ to $t_R$ \cite{Jung,BargerTopAFB,WagnerTopAFB,TaitAFB}.   

Here we extend the maximal flavor violation to the $B$ system and consider the consequences.  Rather than showing that these anomalies could be generated through {\em small} flavor violation between the second and third generations, we consider {\em maximal} flavor violation between the first and third generations.   We will see that the maximal flavor violation in the $B$ system naturally generates an ${\cal O}(20\%)$ contribution to $B_d-\bar{B}_d$ mixing.  The presence of the relatively light right-handed $W',~Z'$s with large couplings to $(u,b)_R$ and $(t,d)_R$ (required for producing the Tevatron forward-backward asymmetry) generates a shift in the $Z$ couplings to right-handed $b$-quarks small enough to be consistent with the LEP measurements of $R_b$ \cite{Choudhury:2001hs}.  
Lastly, since we have chosen the right-handed gauge bosons to couple to only the first and third generations, our model gives rise to the possibility of flavor-changing-neutral-currents.  We discuss the size of the $b \rightarrow s$ mixings necessary to generate the anomalies in $B_s-\bar{B}_s$ mixing and the penguin dominated $b\rightarrow s \bar{q} q$ transitions of $B_d \rightarrow (\phi,\eta',\pi,\rho,\omega) K_s$ \cite{bargerFNUZp}.  Large flavor violation in scalar models \cite{BarShalom:2007pw} and family-nonuniversal $Z'$s \cite{bargerFNUZp,He:2009ie} with small flavor violation have been considered.  

While the concept of large flavor violation has been considered in various contexts before, our goal here is to develop a framework for understanding multiple experimental anomalies that at first sight appear to have very different sources.  Since in our model the flavor violation is part of a gauge structure, this enforces
non-trivial relations between large flavor violation in the top sector and the sensitive flavor observables in the $B$ meson systems. We show that despite strong experimental constraints, the apparently disparate flavor violation in these systems can have the same source.

We begin by fixing the gauge couplings and gauge boson masses through the Tevatron top forward-backward asymmetry.  Di-jet constraints on a $Z'$ gauge boson require its mass $m_{Z'} \gsim 900 \mbox{ GeV}$ \cite{dijet}.  On the other hand, if the gauge boson couplings are to remain perturbative but large enough to generate the top forward-backward asymmetry, we require $m_{W'} \lesssim 600 \mbox{ GeV}$ \cite{Cheung}.  Such a splitting could be generated in principle by adding an additional $U(1)$ (via the same mechanism which splits the $W'$ and $Z'$ in the SM).  However, since the $SU(2)_R$ coupling must be rather large already ($\tilde{g} \simeq 1.5-2$) to generate the forward-backward asymmetry, it is hard to imagine that such an additional coupling from a $U(1)'$ could generate a sufficiently large splitting.  Instead we suppose that a Higgs from a higher representation is responsible for the right-handed $\rho$ parameter deviating from one.  The ratio of the $W'$ and $Z'$ masses is 
$
\frac{m_{W'}^2}{m_{Z'}^2} = \frac{\frac{1}{2}(T(T+1)-T_3^2)}{T_3^2},
$
where $T$ denotes the isospin of the representation, and $T_3$ denotes the isospin component of the Higgs which obtains a vev.  For a triplet Higgs whose $T_3 = -1$ component obtains a vev, $m_{W'}^2/m_{Z'}^2 = 1/2$, which can satisfy the Tevatron dijet constraints for sufficiently heavy $W'$. 
With $m_{W'} \approx 400-600 \mbox{ GeV}$ and $\tilde{g}\approx 1.5-2$, the top $A^{t\bar t}$ can be brought into accordance with experiment while $m_{Z'} \sim \mbox{ TeV}$.  These new gauge bosons give contributions to both single top and top pair production, and will be visible in early LHC data.

\begin{figure}
\includegraphics[width=0.6\linewidth]{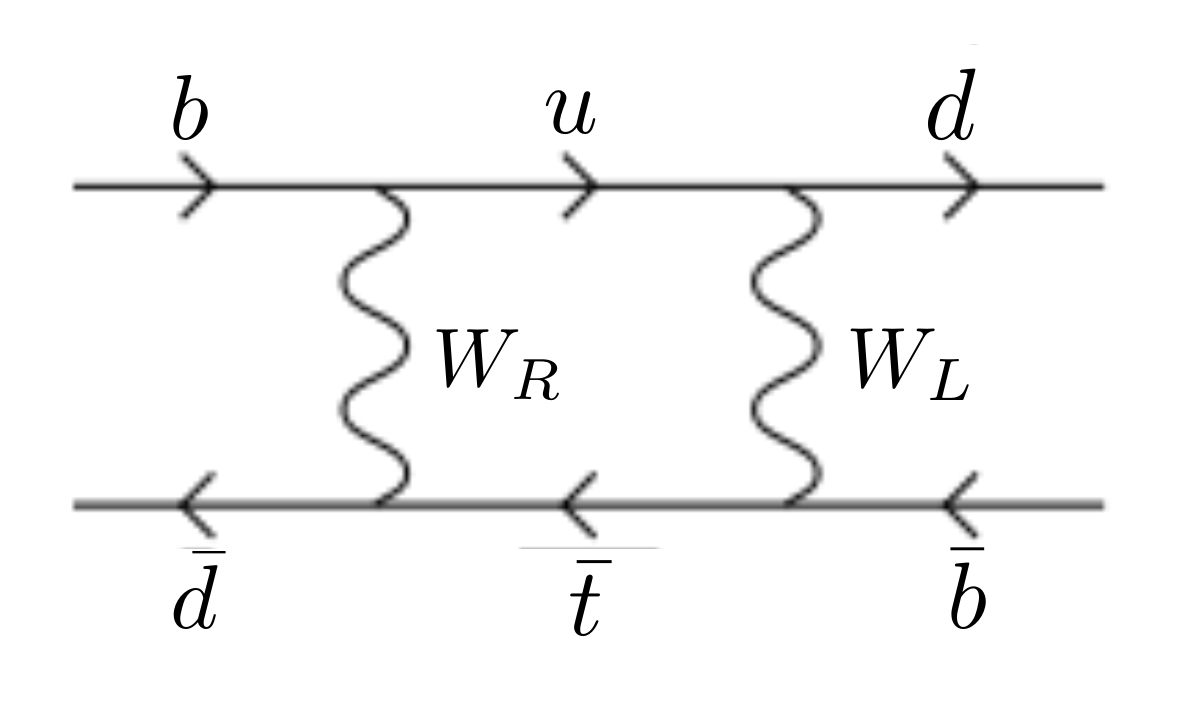}
\caption{Maximal flavor violating contribution to $B_d$ mixing.  Note that there is no CKM suppression at any of the vertices, but the diagram remains small enough due to the chiral suppression of the up quark mass.} \label{fig1}
\end{figure}

Having fixed the masses and couplings for the top forward-backward asymmetry, we consider the effects of coupling $(u,b)_R$ to $W'$.  This gives rise to potentially large contributions to $B_d-\bar{B}_d$ mixing, through the diagram shown in Fig.~(\ref{fig1}).  Na\"{\i}vely these contributions would be disastrously large, because there is no CKM suppression at the vertices.  There is however a chiral suppression of the light quark mass $m_u$ which makes this contribution phenomenologically feasible: 
\begin{eqnarray}
{\cal M}_{LR} &=& -g^2 \tilde{g}^2\frac{m_t m_u}{32 \pi^2 m_{W'}^4}{\cal O}_{LR} \nonumber \\ 
&& \left(\frac{\log \beta}{(1-\beta)(x'_t-\beta)}+\frac{\log x'_t}{(1-x'_t)(\beta-x'_t)}   \right),
\end{eqnarray}
where $x'_t = m_t^2/m_{W'}^2$, $\beta = m_W^ 2/m_{W'}^ 2$, and ${\cal O}_{LR} = \left[\bar{u}^\alpha_d \frac{1}{2}\gamma_\nu(1-\gamma_5) u_b^\alpha\right] \times \left[\bar{v}^\beta_d \frac{1}{2}\gamma_\nu(1+\gamma_5) v_b^\beta\right].$
Here we have used the computation of \cite{Beall}.
This is to be compared against the Standard Model results:
\begin{equation}
{\cal M}_{LL} \simeq g^4 V_{td}^2 \eta_t H(x_t) \frac{m_t^2}{32 \pi^2 m_W^4} {\cal O}_{LL},
\end{equation} 
where $H(x_t) = \frac{1}{4}+\frac{9}{4}\frac{1}{1-x_t}-\frac{3}{2}\frac{1}{(1-x_t)^2}-\frac{3}{2}\frac{x_t^2}{(1-x_t)^3}\log x_t \simeq 0.55$
and $ \eta_t \approx 0.58$ accounts for QCD corrections.
The matrix elements are read off in the standard way: 
\begin{eqnarray}
\langle \bar{B}_d | {\cal O}_{LL} | B_d \rangle & =& B_d^{(1)} \frac{1}{3} f_{B_d}^2 m_{B_d} \nonumber \\
\langle \bar{B}_d | {\cal O}_{LR} | B_d \rangle & =&  B_d^{(4)} \frac{ m_{B_d}^3}{4(m_b+m_d)^2} f_{B_d}^2.
\end{eqnarray}
with $B_d^{(1)} = 0.8$ and $B_d^{(4)} = 1.16$. The ratio of the amplitudes is
\begin{eqnarray}
\frac{{\cal M}_{LR}}{{\cal M}_{LL}} \simeq -0.2\left(\frac{\tilde{g}}{2} \right)^2 \left(\frac{450 \mbox{ GeV}}{m_{W'}} \right)^4 \frac{G(x_t)}{G(x_t^0)},
\end{eqnarray}
where we have used $m_u = 2.5 \mbox{ MeV}$, $V_{td} = 8 \times 10^{-3}$, and $G(x_t) = (\log\beta/(1-\beta)-\log x_t'/(1-x_t'))/(x_t'-\beta)$, where $G(x_t^0)$ is evaluated at $m_{W'} = 450 \mbox{ GeV}$.  Such an amplitude of $B_d - \bar{B}_d$ mixing is consistent with the size of the new physics contributions in a global fit \cite{Ligeti} to the $B_s$ and $B_d$ anomalies described in the introduction, in concert with the new contributions to $B_s - \bar{B}_s$ mixing described below.  

A new CP-violating phase $\phi_d^0$ could also contribute to the discrepancy in the measurement of $\sin 2\beta$ in $B_d \rightarrow \psi K_S$ with respect to the value obtained from CKM triangle fits.    Using a definition $S_{\psi K} = \sin\left[2\beta + \mbox{arg}(1+ h_d e^{2 i \phi_d})\right]$, an angle $\phi_d$ between  $\pi/2$ and $3 \pi/4$ could contribute to $-\eta_{CP} S_{\psi K} = 0.655 \pm 0.0244 $ to explain the discrepancy between it and the best-fit CKM triangle value obtained in \cite{Lunghi:2010gv}: $\sin(2\beta)^{\mbox{fit}}= 0.891 \pm 0.052$.  

By itself, $B_d$ mixing is not sufficient to generate the Tevatron di-muon asymmetry.   However, the non-universality of the $W'$ and $Z'$ couplings provides a mechanism for generating flavor changing neutral currents. In particular, a slight misalignment of mass and flavor bases for the $b_R $ and $s_R $ quarks will give rise to $b$--$s$ flavor-changing neutral currents which provide a new contribution to $B_s$ mixing.  Compatibility with the Tevatron data then determine the magnitude and phase of the $b$--$s$ mixing angle, $S_{sb}$.  The flavor-changing operator induced by the mass-flavor misalignment is $ \frac{S_{sb}^2 \tilde{g}^2}{m_{Z'}^2} \; \bar{b}_R \gamma^\mu s_R \bar{b}_R \gamma_\mu s_R .$
The effects of flavor-changing $b \rightarrow s$ transitions have already been analyzed in the literature, and we make use of the result here \cite{Dobrescu}.  In order to generate an asymmetry of comparable size to that at the Tevatron we require
\begin{equation}
\Lambda \equiv \frac{m_Z'}{|S_{sb}| \tilde{g}} \sim 0.4-1.2 \mbox{ TeV}\left(\frac{10^{-2.5}}{\tilde{g} |S_{sb}|}\right),
\end{equation}
giving an approximate size to the flavor violation.  One might worry that these flavor changing processes could contribute to $B_s \rightarrow \mu^+\mu^-$ processes since $Z'$ couples to hypercharge and hence picks up a coupling to $\mu^+ \mu^-$ of size $g_1^2/\tilde{g}$, where $g_1$ is the hypercharge coupling.  The branching fraction for this process is 
\begin{eqnarray}
{\cal B}(B_s \rightarrow \mu^+ \mu^-) & = &  \frac{|S_{bs}|^2 g_b^2 g_\mu^2 (y_{\mu L}^2 + y_{\mu R}^2)}{m_{Z'}^4} \frac{\tau_{B_s} M_{B_s}m_\mu^2 f_{B_s}^2}{64 \pi} \nonumber \\
& \approx & 10^{-12} \left(\frac{|S_{sb}|}{10^{-2.5}}\right)^2\left(\frac{1 \mbox{ TeV}}{m_{Z'}}\right)^4
\end{eqnarray}
where $g_\mu = g_1^2/\tilde{g}$ and $g_b = -\frac{1}{2} \tilde{g}$.  This is well below the current limit of $4.3 \times 10^{-8}$ \cite{CDFBs}.

We also note that such flavor violating decays may give contributions to the penguin dominated decays of $b \rightarrow s \bar{q} q$ in the processes $B_d \rightarrow (\phi,\eta',\pi,\rho,\omega,f_0)K_s$.   The new contributions may further serve to suppress $S_{fCP}$ relative to the tree-level decay in $S_{\psi K}$, as explored in \cite{Barger:2009qs}.  Here we simply wish to note that the $Z'$-mediated interaction is of the right magnitude to contribute to $S_{fCP}$ for the interactions $B_d \rightarrow (\phi,\eta',\pi,\rho,\omega,f_0)K_s$.  In particular the combination $\Delta C_{NP} = (C_{uu}^{bs} - C_{dd}^{bs})/V_{tb} V_{ts}$
must be ${\cal O}(1)$ (or slightly smaller) in order to give rise to the correct shifts in $S_{fCP}$ \cite{Barger:2009qs}.  Here $C_{qq}^{bs} = S_{sb} T_{3R,q}(\tilde{g}\,m_Z/g\, m_{Z'})^2$.  We find with $S_{sb} = 10^{-2.5}$, $m_{Z'} = 900 \mbox{ GeV}$, $\tilde{g} = 2$, that the $b \rightarrow s$ transitions may contribute to the penguin dominated $B_d$ processes.  Measurements of $S_{fCP}$ in these modes at future $B$ factories could therefore provide a detailed test of this model.

Next we calculate the size of the shift of the right-handed coupling of the $Z$ to the $b$ quarks, which affects the $Z$ pole observables. 
The symmetry-breaking pattern is $SU(2)_L \times SU(2)_R \times U(1)' \rightarrow U(1)_Y \times SU(2)_L$.  
The mass matrix, in the $(W_L^3,B,W_R^3)$ basis is
\begin{equation}
{\cal M} = \frac{1}{4}
\left(
\begin{array}{ccc}
g^2 v^2 &  g g' v^2 & 0 \\
g g' v^2 & {g'}^2 (v^2 + 4 Y^ 2 \hat{v}^2) & 4 Y T_{3R} g'\tilde{g} \hat{v}^2 \\
0 &  4 YT_{3 R} g'\tilde{g} \hat{v}^2 &  4 T_{3 R} ^ 2 \tilde{g}^2 \hat{v}^2
\end{array}
\right),
\end{equation}
where $\hat{v}$ is the right-handed Higgs vev, $g'$ is the $U(1)'$ coupling, and we will henceforth take the right-handed Higgs hypercharge $Y$ equal to its isospin $T_{3R}$ for simplicity.  By diagonalizing the mass matrix, in the $(W_L^3,B,W_R^3)$ basis, the $Z$ mass eigenstate has components
\begin{equation}
Z = 
\left(
\begin{array}{c}
\frac{g}{g_Z} \\
\frac{-g'\tilde{g}^2}{(\tilde{g}^2+g'^2)g_Z}(1+\epsilon_Z^2 \frac{{g'}^2 g_Z^2}{g^2(\tilde{g}^2+g'^2)}) \\
\frac{-g'^2\tilde{g}}{(\tilde{g}^2+g'^2)g_Z}(1+\epsilon_Z^2 \frac{\tilde{g}^2 g_Z^2}{g^2 (\tilde{g}^2+g'^2)})
\end{array}
\right),
\end{equation}
where $g$ is the SU(2)$_L$ coupling, and we have expanded the expression to second order in $\epsilon_Z = m_Z/m_{Z'}$ and
$
g_Z = \sqrt{\frac{(\tilde{g}^2+g'^2)g^2+g'^2\tilde{g}^2}{(\tilde{g}^2+g'^2)}},
$
so that we can make the identification of the $U(1)_Y$ coupling as
$
g_1^2 = \frac{g'^2\tilde{g}^2}{(\tilde{g}^2+g'^2)}.
$
Thus we can read off the couplings of left- and right-handed fields to the $Z$ boson:
\begin{equation}
g_L = g_Z(T_3^L - s_W^2 Q) + \epsilon_Z^2 (Q-T_3^L) g_Z \frac{g_1^4}{\tilde{g}^2 g^2}
\end{equation}
and
\begin{equation}
g_R = -Q \frac{g_1^2}{g_Z}  - \epsilon_Z^2 (\frac{g_1^4}{\tilde{g}^2 g^2}(T_3^R-Q)+\frac{g_1^4}{g'^2 g^2} T_3^R) g_Z.
\end{equation}
As a representative choice, we take $\tilde{g} = 2$, $m_{Z'} = 900 \mbox{ GeV}$, in which case we find $\delta g_b^R \simeq 0.001$ and $\delta g_b^L  \simeq 10^{-5}$.  These shifts are too small to give substantial impact to the measurement of $R_b$ and $A_{FB}^b$ at LEP \cite{Choudhury:2001hs}.  Contributions to the $\rho$ parameter
are present at tree level, 
\begin{equation}
\frac{\delta \rho}{\rho}=-\frac{\delta m_Z ^ 2}{m_Z ^ 2} =\frac{ (m ^ 2_Z-m_W ^ 2)\tan ^ 2\theta_R}{m^2_{Z'}- (m ^ 2_Z-m_W ^ 2)\tan ^ 2\theta_R},
\end{equation}
where $\sin\theta_R =g'/\sqrt{g'^2+\tilde{g}^2}$ is the right-handed analogue of the Weinberg angle.  
For $g_R=2$, $m_{Z'}= 900\gev$, this contribution is less than one part in $10^{4}$, safely within experimental bounds.
Contributions to $S$ are negligible, as the absence of any new fields transforming under $SU(2)_L$ means that
the leading contributions to $W_L^3-B$ mixing are precisely those in the SM.


Lastly, we discuss specifics of the mass and flavor model to ensure that something reasonable can be written down. In the $SU(2)_L$ flavor basis for left-handed quarks $q'_L$ and the $SU(2)_R$ flavor basis for right-handed quarks $q'_R$, the gauge boson interactions take the form 
\begin{equation}
{\cal L}= {q'}_L \tau^a \gamma^\mu {q'}_L W_\mu^a + q'_R F' \tau^a \gamma^\mu q'_R {W'}_\mu^a,
\label{SU2basisL}
\end{equation}
where $F'$, a coupling matrix, is being left explicitly in the expression to account for the non-universal coupling of the quarks to the right-handed fields. 
Now rotating to the mass basis, we have
\begin{eqnarray}
{\cal L} &=&  
\bar{d}_L V_{CKM} \gamma^\mu u_L W^-_\mu +  \bar{d}_L  \gamma^\mu d_L Z_\mu +  \bar{u}_L  \gamma^\mu u_L Z_\mu \nonumber \\
&& +  \bar{d}_R \tilde{V}_{CKM} \gamma^\mu u_R {W'}^-_\mu +  \bar{d}_R F_d  \gamma^\mu d_R Z_\mu  \nonumber \\
&& +  \bar{u}_R  F_u \gamma^\mu u_R Z_\mu,
\end{eqnarray}
where the relation between ${q'}_{L,R}$ and the mass basis quarks $q_{L,R}$ is
\begin{eqnarray}
d_L = S_d d'_L,~~u_R = S_u u'_L \nonumber \\
d_R = S'_ d d'_R,~~u_R = S'_u u'_R
\end{eqnarray}
with
\begin{eqnarray}
V_{CKM} = S_d^\dagger S_u,~~~\tilde{V}_{CKM} =  {S'}_d^\dagger F' {S'}_u \nonumber \\
F_u = {S'}_u^\dagger F' {S'}_u,~~~F_d = {S'}_d^\dagger F' {S'}_d.
\end{eqnarray}
Complex phases in the right-handed flavor mixing matrices are physical, as setting the form of the left-handed CKM matrix uses the entire rephasing freedom in the SM.  With the nonuniversal couplings of $SU(2)_R$, $F' = \mbox{Diag}(1,0,1)$, and FCNC's are generated at tree level.
We now briefly examine a consistent choice of $S'_u$ and $S'_d$ that generate the $b \rightarrow s$ FCNC while simultaneously maintaining the form of the right-handed CKM matrix which pair $(u,b)_R$ and $(t,d)_R$ in their couplings to the right-handed fields.  Consider for example the pair of matrices:
\begin{equation}
S'_u = 
\left(
\begin{array}{ccc}
0 & 0 & -1 \\
0 & 1 & 0 \\
1 & 0 & 0
\end{array}
\right),~~~~~~~
S'_d \approx 
\left(
\begin{array}{ccc}
1 & 0 & 0 \\
0 & 1 & -\epsilon \\
0 & \epsilon & 1
\end{array}
\right),
\end{equation}
where $\epsilon$ generates the flavor violating $b \rightarrow s$ processes.  This gives rise to a right-handed CKM matrix
\begin{equation}
\tilde{V}_{CKM} \approx 
\left(
\begin{array}{ccc}
0 & 0 & -1 \\
\epsilon & 0 & 0 \\
1 & 0 & 0
\end{array}
\right),
\end{equation}
implying a small contribution in $u \rightarrow s$.  This contributes to Kaon mixing through a left-right diagram of the type in Fig.~(1), but with $\bar{d} s$ on external lines.  This diagram scales as $g^2 \tilde{g}^2 m_t m_u \epsilon V_{ts}/(m_{W'}^2 m_W^2)$ which is to be compared against the SM contribution which scales as $g^4 V_{cs}^2 V_{cd}^2 m_c^2/m_W^4$.  The new contribution is suppressed relative to the SM by approximately 1 part in $10^4$, well below the experimental uncertainty in the Kaon mass difference.
Meanwhile, the FCNC induced in down-type quarks is of the desired form
\begin{equation}
F_d \approx 
\left(
\begin{array}{ccc}
1 & 0 & 0 \\
0 & 0 & \epsilon \\
0 & \epsilon & 1
\end{array}
\right).
\end{equation}

With couplings only to right-handed quark doublets, the $SU(2)_R $ is anomalous, which could be remedied by adding new heavy fermions.  
Since there are many possibilities for anomaly cancellation that give rise to the same phenomenology that we study here, we do not consider it further.  Mass generation is a more complicated issue.  The first and third generation right-handed fields have a Higgs term
\begin{equation}
\frac{Y^d_{ij}}{M}{\bar{q'}}_R^i \phi_R^\dagger H_L {q'}_L^{j} + \frac{Y^u_{ij}}{M}{\bar{q'}}_R^i \tilde{\phi}_R^\dagger \tilde{H}_L {q'}_L^{j} ,
\label{higherdimensionhiggs}
\end{equation}
where $i = 1,3$ and $j = 1,2,3$, whereas the second generation right-handed fields take on the more conventional form
\begin{equation}
Y_d^j{\bar{q'}}_R^2  H_L {q'}_L^{j} + Y_u^j{\bar{q'}}_R^2  \tilde{H}_L {q'}_L^{j}.
\end{equation}
Here we have introduced a new $SU(2)_R $ doublet $\phi_R$ which gives a secondary contribution to $SU(2)_R $ symmetry breaking, and $\tilde{H}_L^ a=\epsilon_{ab}H_L ^*,\tilde{\phi}_R ^ a =\epsilon_{ab}\phi ^*_R $ are the conjugate representations.  This method for mass generation
does not address the common difficulty of generating the large top Yukawa coupling.
Note that we do not make use of a left-right Higgs for the couplings in Eq.~(\ref{higherdimensionhiggs}), as this gives rise to mixing between the left and right-handed gauge bosons, inducing dangerous contributions to, {\em e.g.}, $B \rightarrow \tau \nu$.  The higher dimension Higgs couplings of Eq.~(\ref{higherdimensionhiggs}) can be generated by integrating out $SU(2)_{L,R}$-singlet fermions $T^i$, $B^i$ which carry the appropriate hypercharge.  
These quarks have vector-like masses, $m_{ij}\bar T^i T^j$, and
upon integrating them out the SM Yukawa couplings are generated with $Y^{ij}/M = y_R^{ik}m^{-1}_{kl}y_L^{lj}$.  The flavor structure of the SM
is then generated from an enlarged flavor structure including vector-like quarks, together with hierarchies
from seesaw-like mechanisms operating in the massive quark sector.  The different flavor structures of
left- and right-handed quarks in our model then naturally generate different mixing structures in 
their respective CKM matrices.  The generation of the SM Yukawa couplings from a larger flavor structure at relatively light mass scales $M$ yields a new approach to generating flavor hierarchies in the quark sector, a full exploration of which we leave for future work.

To conclude, we considered a model of maximal flavor violation connecting the first and third generation quarks.  We showed that the flavor physics which generates the Tevatron top forward-backward asymmetry may be closely connected to apparently much smaller flavor violation in the $B_d$ and $B_s$ systems.  In particular, maximal flavor violation connecting only the first and third generations naturally gives rise to SM size contributions to $B_d$ mixing on account of the chiral  $m_u$ suppression of the new left-right CKM diagrams, despite the absence of CKM suppression in the vertices.  $B_s$ mixing arises through small flavor-changing-neutral-currents arising from $b-s$ mass mixing in the presence of a non-universal $Z'$.  Our model predicts new physics in $B_d$ mixing, and for the reference parameter values chosen here, predicts a contribution to the semileptonic asymmetry $A^d_{sl} \lsim -0.002$, which may be verified in asymmetric $B$ factories.

While the concordance of all these anomalies within the context of a relatively simple model is intriguing, a more general lesson to draw from this model is that flavor violation in new physics explaining $B$ meson observables need not be small.  It is also easy to connect this large flavor violation to large flavor violation in the top system.  Where the mixing involves the first and third generation, maximal flavor violation operns new possibilities for model building.

{\em Acknowledgements:}  We thank C.~Cheung, B.~Dobrescu, M.~Gresham, J.~Hewett, I.-W.~Kim, Z.~Ligeti, Y.~Nomura, A.~Pierce, W.~Skiba, T.~Volansky, and H.-B.~Yu for discussions.  JS was supported in part by DOE grant DE-FG02-92ER40704.

{\em Note Added:} This model makes several distinctive predictions for the early LHC.  The $W'$ will be visible as a resonance in top-light jet as little as $1 \mathrm{fb}^{-1}$ \cite{Gresham:2011dg}, while new contributions to single top production through $ud\to t b$ could lead to distinctive signals even within $\lsim 1\mathrm{fb}^{-1}$ \cite{Craig:2011an}.

\end{document}